\documentclass[%
 aip,
 jap,%
 amsmath,amssymb,
reprint,%
]{revtex4-1}

\usepackage{graphicx}
\usepackage{dcolumn}
\usepackage{bm}
\usepackage[caption=false]{subfig}

\begin{document}

\preprint{AIP/123-QED}

\title[Sputter gas pressure effects on the properties of Sm-Co thin films deposited from a single target.]
{Sputter gas pressure effects on the properties of Sm-Co thin films deposited from a single target.}

\author{T.~G.~A.~ Verhagen, D.~B.~ Boltje, J.~M.~ van Ruitenbeek and J.~ Aarts}
\email{aarts@physics.leidenuniv.nl}
\affiliation{Kamerlingh Onnes Laboratorium, Universiteit Leiden, PO Box 9504, 2300 RA Leiden, The Netherlands}

\date{\today}

\begin{abstract}
We grow epitaxial Sm-Co thin films by sputter deposition from an alloy target
with a nominal SmCo$_5$ composition on Cr(100)-buffered MgO(100) single-crystal
substrates. By varying the Ar gas pressure, we can change the composition
of the film from a SmCo$_5$-like to a Sm$_2$Co$_7$-like phase. The composition,
crystal structure, morphology and magnetic properties of these
films have been determined using Rutherford Backscattering, X-ray
diffraction and magnetization measurements. We find that the
various properties are sensitive to the sputter background pressure in
different ways. In particular, the lattice parameter changes in a continuous
way, the coercive fields vary continuously with a maximum value of 3.3~T, but the saturation magnetization peaks
when the lattice parameter is close to that of Sm$_2$Co$_7$.  Moreover, we find that the Sm content of the films 
is higher than expected from the expected stoichiometry.
\end{abstract}


\maketitle

\section{Introduction}
Modern permanent magnetic materials, like SmCo$_5$ and NdFe$_{14}$B, are based
on intermetallic compounds of rare-earth and 3d transition metals. Sm-Co
intermetallics are hard magnetic materials with a high coercive field and a
high uniaxial magnetocrystaline anisotropy, where the easy axis is aligned
along the crystallographic $c$-axis. Since the 1970s/1980s many groups
investigated the properties of Sm-Co crystals and thin films. The control of
the composition and the crystallographic texture are the key parameters to
obtain thin films with the desired hard magnetic properties. These properties
are interesting from both a technical and a fundamental point of view. The
further miniaturization of magnetic microelectromechanical systems (MEMS)
\cite{Chin200075} requires the control of such films. But also the combination
between a hard magnet like SmCo$_5$ with soft magnets \cite{sawatzki:123922} or
superconductors \cite{Engelmann} is an unexplored area that can lead to
interesting and useful magnetic configurations.

In the last years, recipes have been developed to grow films with the desired
hard magnetic properties. One way to obtain them is to grow epitaxial
thin films. Epitaxial growth can be obtained by using MgO(100), MgO(110) and
Si(100) single crystals, commonly in combination with a chromium buffer layer.
Growing Sm-Co films on the four-fold symmetric MgO(100) substrates results in
the epitaxial relation Sm-Co(11-20)[0001]//Cr(001)[110]//MgO(001)[100],
where in the film the Sm-Co grains are equally distributed along the two
in-plane directions. Sm-Co films can also be deposited on a glass substrate.
Growing on glass results in very small crystallites in a disordered
structure, and yields large coercive fields \cite{zhang:113908}.

Most groups grow thin Sm-Co films using pulsed-laser deposition (PLD)
\cite{singh:072505} or sputter deposition \cite{fullerton:2438} from single
elemental targets Sm and Co. By tuning the sputter power of both sources or the
pulse ratio by PLD, it is possible to grow compositions in the desired
range. Also the influence of the substrate temperature has been studied,
for obtaining epitaxial films. The effect on the film growth of the sputter
background pressure is often not taken into account. Still, the sputter
pressure plays an important role in the growth kinetics and one recent
study showed that changes in the stoichiometry and magnetic properties occur
when varying the pressure \cite{Speliotis20051195}. One of the underlying
problems is the complexity of the Sm-Co phase diagram
\cite{springerlink:10.1361/105497199770340815} in which, on the Co-rich side,
the compounds Sm$_2$Co$_{17}$ (11~\% Sm), SmCo$_{5}$ (17~\% Sm),
Sm$_5$Co$_{19}$ (21~\% Sm) and Sm$_2$Co$_{7}$ (22~\% Sm) all exist; with the note
that SmCo$_5$ actually is a metastable compound. The connection between
composition and magnetic properties is therefore not trivial. Fortunately, high
coercive fields can be found over a range of compositions and a number of
studies has focused on this particular aspect. \\
In this paper, we show that by varying the argon pressure, we are able to
grow SmCo$_5$-like and Sm$_2$Co$_7$-like phases from an alloy target, with good
crystallographic texture, coercive field and saturation magnetization. The
properties of the grown films are shown to be very sensitive to the sputter
pressure used.

\begin{figure*}
    \includegraphics[width=2\columnwidth]{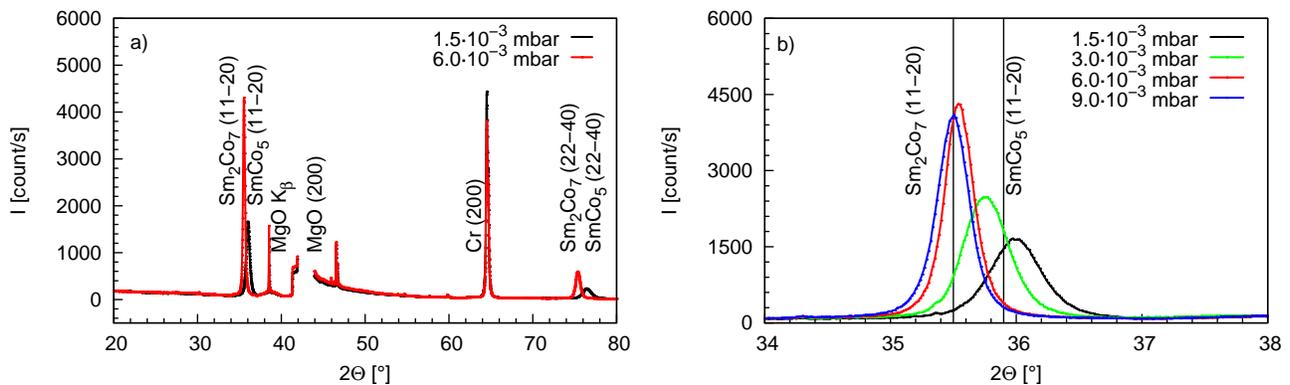}\\
\caption{a) $\theta$-2$\theta$ XRD scans of Sm-Co films grown from an Sm-Co
alloy target on a Cr/MgO(100) substrate with two different sputter gas pressures
as indicated. b) shows the region around the Sm-Co(11-20) peak, where the
two vertical lines indicate the reflection of bulk crystalline Sm$_2$Co$_7$
(left) and SmCo$_5$ (right). Four samples are shown, grown at (from right
lo left), 1.5 (black), 3.0 (green), 6.0 (red) and 9.0 $\cdot$~10$^{-3}$ mbar
(blue).} \label{fig:figure_1}
\end{figure*}

\section{Experiment}
The Sm-Co films were deposited in a UHV chamber (base pressure 1~$\cdot$~
10~$^{-9}$ mbar) using DC magnetron sputter deposition with argon as plasma
from a commercially obtained alloy target with a nominal composition of
SmCo$_5$ (3N). A rotating sample holder was used. The deposition rate was
measured by X-ray reflectivity (XRR) using Cu-K$\alpha$ radiation.

Films were deposited on 500 $\mu$m thick MgO(100) single crystal substrates
on which a 100 nm thick Cr buffer layer was first deposited at
250$^{\circ}$C in an Ar pressure of 1.5~$\cdot$~10$^{-3}$~mbar. All Sm-Co films
were approximately 100 nm thick and were grown at 450$^{\circ}$C with an
Ar pressure varying between 1.5~$\cdot$~10$^{-3}$~mbar and
12.5~$\cdot$~10$^{-3}$~mbar. Afterwards, a 10 nm thick Cr layer was
deposited at 450$^{\circ}$C as a protection layer.

The actual film composition and thickness were determined using Rutherford
Backscattering (RBS). The structural quality of the film was measured with
$\theta$-$2\theta$ X-ray diffractometry (XRD) using Cu-K$\alpha$ radiation,
where the MgO substrate peak was measured as a reference for the angle,
by using an extra Cu-absorber to decrease the intensity. The morphology of the
films was characterized by Atomic Force Microscopy (AFM) in tapping mode.
Magnetization measurements were performed in a SQUID-based magnetometer
(MPMS 5S from Quantum Design) in fields up to 5 T. For the magnetization
measurements, the substrates were cut in pieces of approximately 10~$\times$~4~
mm$^2$. As a reference, an MgO(100) substrate was measured, and also 
an MgO(100) substrate with a 100 nm Cr film protected with 30 nm Cu.
Electron paramagnetic resonance (EPR) spectra were measured at room temperature using a Bruker
EMX plus X-band spectrometer in a TE$_{011}$ cavity with 100~kHz modulation frequency and 1~G modulation amplitude.

\section{Results}
\begin{figure}
\includegraphics[width=\columnwidth]{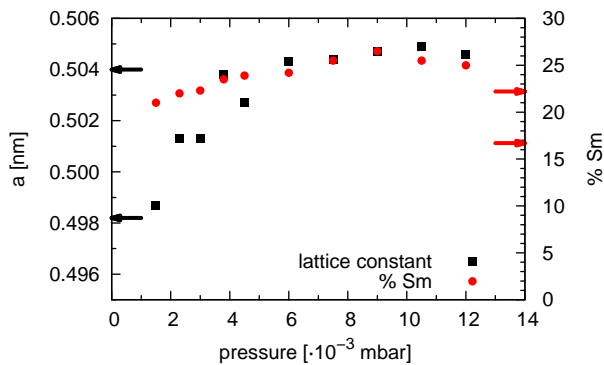}
\caption{The lattice constant and the Sm concentration of the Sm in Sm-Co films
as a function of the sputter background pressure, where the black ($\leftarrow$) and red ($\rightarrow$) arrows indicate the lattice constant and Sm concentration of bulk SmCo$_5$ and Sm$_2$Co$_7$. 
Clearly visible is the change of the lattice parameter a from the Sm$_2$Co$_7$ phase (a=0.5040~nm) to the
SmCo$_5$ phase (a=0.4982~nm).} \label{fig:figure_2}
\end{figure}

Figure~\ref{fig:figure_1}(a) shows the XRD scans of films grown at 1.5~$
\cdot$~10$^{-3}$~mbar (SmCo$_5$-like) and 6.0~$\cdot$~$10^{-3}$~mbar
(Sm$_2$Co$_7$-like) respectively. The observed peaks are labeled as reflections
of Sm-Co, MgO and Cr. Due to the thickness and high crystallinity, also
the K$_{\beta}$ peak of the MgO substrate is visible. In Figure~\ref{fig:figure_1}(b)
the region around the Sm-Co(11-20) peak
is shown, for films grown with a sputter pressure of 1.5, 3.0, 6.0 and 9.0~
$\cdot 10^{-3}$~mbar. Clearly visible is that, with decreasing pressure, the
peaks shift to a higher angle. The measured lattice constant, determined from the Sm-Co(22-40) peak,
and the Sm content are plotted in Figure~\ref{fig:figure_2} as a function of the
sputter pressure. For films grown at a pressure above 6.0~$\cdot$~10$^{-3}$~mbar,
the lattice parameter of the Sm-Co film is almost that of bulk
Sm$_2$Co$_7$ (0.5040~nm). Decreasing the pressure from 6.0~$\cdot$~
10$^{-3}$~mbar shows a decreasing lattice parameter, and at the lowest pressure
the lattice constant of the Sm-Co film just reaches the SmCo$_5$ bulk value
(0.4982~nm). With respect to the Sm concentration, we consistently find a
somewhat higher number than the stoichiometric Sm-Co phases would yield. Above
6.0~$\cdot$~10$^{-3}$~mbar, the Sm concentration is around 25~-~27~\%
(compared to 22~\% for Sm$_2$Co$_7$). Below 6.0~$\cdot$~10$^{-3}$~mbar the Sm
content gradually decreases to 21~\% (compared to 17~\% for SmCo$_5$).

\begin{figure}%
    \includegraphics[width=\columnwidth]{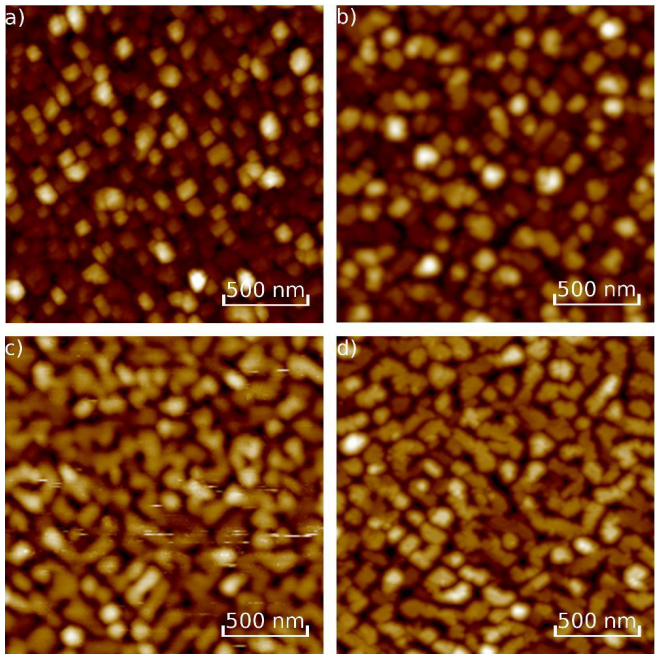}
\caption[]{Morphology of the Sm-Co film grown at a) $1.5 \cdot 10^{-3}$, b) $3
\cdot 10^{-3}$, c) $6 \cdot 10^{-3}$  and d) $9.0 \cdot 10^{-3}$ mbar measured
with atomic force micropscopy. } \label{fig:figure_3}
\end{figure}

Figure~\ref{fig:figure_3}(a)-(d) show the morphology of the Sm-Co films grown at
1.5, 3.0, 6.0 and 9.0 $\cdot$~10$^{-3}$~mbar, respectively. The films grown at a
high sputter background pressures consists of rectangular grains with an
average size of 70~$\times$~250~nm$^2$. Statistical analysis over an area of 1~
$\times$~1~$\mu$m$^2$ on the Sm-Co film grown at 6~$\cdot$~10$^{-3}$ mbar
indicates an average surface roughness of 8.1 nm and a peak to peak value of
57~nm. When decreasing the pressure below 6~$\cdot$~10$^{-3}$~mbar, the shape of
the grains slowly transforms from rectangular to square-like. Sm-Co films
grown at 1.5~$\cdot$~10$^{-3}$~mbar consist of square grains with an average size
of 75~$\times$~75~nm$^2$. Statistical analysis over an area of 1~$\times$~1~
$\mu$m$^2$ on the Sm-Co film grown at 1.5~$\cdot$~10$^{-3}$~mbar indicates an
average surface roughness of 9.6 nm and a peak to peak value 55 nm.

\begin{figure}%
    \includegraphics[width=\columnwidth]{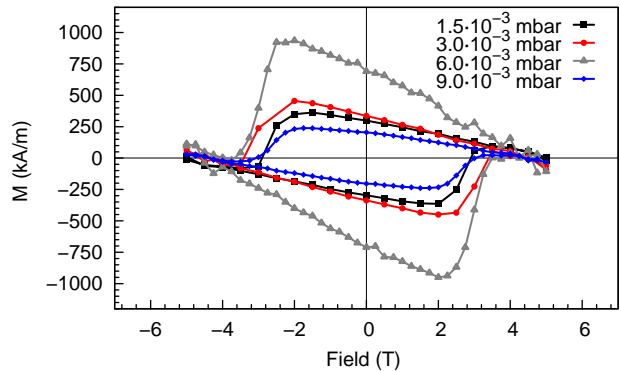}%
    \caption[]{Uncorrected magnetic hysteresis of the Sm-Co film on a Cr/MgO(100)
 substrate grown with different sputter gas pressures as indicated at 300~K.}
\label{fig:figure_4}%
\end{figure}

In Figure~\ref{fig:figure_4} the magnetization measurements are shown,
taken at room temperature. The magnetization was calculated by dividing the
measured magnetic moment by the measured volume of the Sm-Co films (typically
10 mm $\times$ 4 mm $\times$ 100 nm). All samples show hysteretic
behavior with a square-like loop and a large coercivity, of the order of
3~T, but also a substantial diamagnetic contribution. Separate measurements on
an MgO(100) substrate and an MgO(100)/Cr(100 nm)/Cu(30 nm) film show that, 
at room temperature, the magnetic susceptibility $\chi$ of MgO for
substrates from different batches varied and the samples measured had a
magnetic susceptibility of -2.4 and -3.5~$\cdot$~10$^{-7}$~emu/g. These values are in agreement with
$\chi$ = -5.1~$\cdot$~10$^{-7}$~emu/g for a single-crystal MgO \cite{Gordon}.

\begin{figure}
\includegraphics[width=\columnwidth]{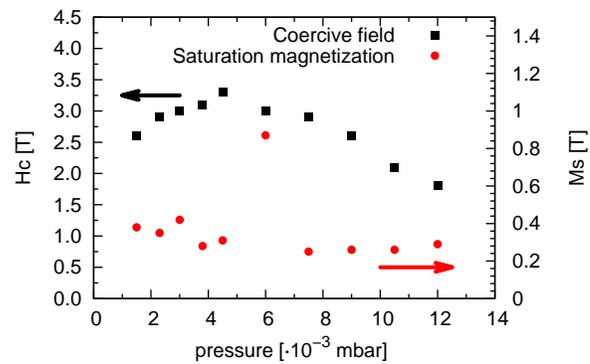}
\caption{The coercive field and the saturation magnetization of the Sm-Co films
as a function of the sputter background pressure.} \label{fig:figure_5}
\end{figure}

For the films grown at 1.5, 3.0 and 9.0~$\cdot$~10$^{-3}$~mbar, the coercive
field $H_c$ is approximately 2.5~T and the saturation magnetization $M_s$ is
approximately $0.4$ T. The film grown at 6.0~$\cdot$~10$^{-3}$~mbar has a
slightly higher coercive field of 3.0~T and a significantly larger
saturation magnetization of 0.87~T. Both $H_c$ and $M_s$ as a function of
sputter pressure are given in Figure~\ref{fig:figure_5}. Again we find
clear trends: with increasing pressure up to 6~$\cdot$~10$^{-3}$~mbar $H_c$
slowly increases from 2.6~T to 3.3~T, but above 6~$\cdot$~10$^{-3}$~mbar a rapid
decrease sets in, down to 1.8~T at 12~$\cdot$~10$^{-3}$~mbar. The saturation
magnetization is 0.3-0.4~T in the whole pressure range, but clearly visible is
the much higher saturation magnetization of the film grown at 6.0~$\cdot$~
10$^{-3}$ mbar. To characterize the magnetic texture, the ratio of the remnance
of the in-plane and out-of-plane magnetizations was determined for the
films grown at 1.5 and 6.0~$\cdot$~10$^{-3}$~mbar (not shown). These data
show that both films have a preferred in-plane texture.

\begin{figure}
    \includegraphics[width=\columnwidth]{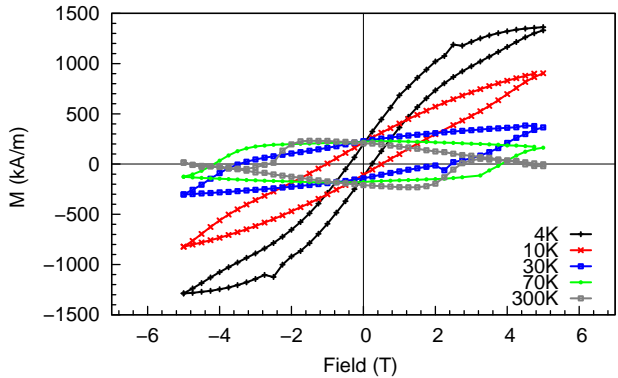}
\caption[]{a) Uncorrected magnetic hysteresis of the Sm-Co film grown at 10.5~$\cdot$~10$^{-3}$~mbar
as a function of temperature. For the 10~K and 4~K loop, the 5~T magnet of the
magnetometer is not strong enough to reach the coercive field of the Sm-Co film
and the contribution of the MgO substrate increases significantly.}
\label{fig:figure_6}%
\end{figure}

\begin{figure*}%
	\includegraphics[width=2\columnwidth]{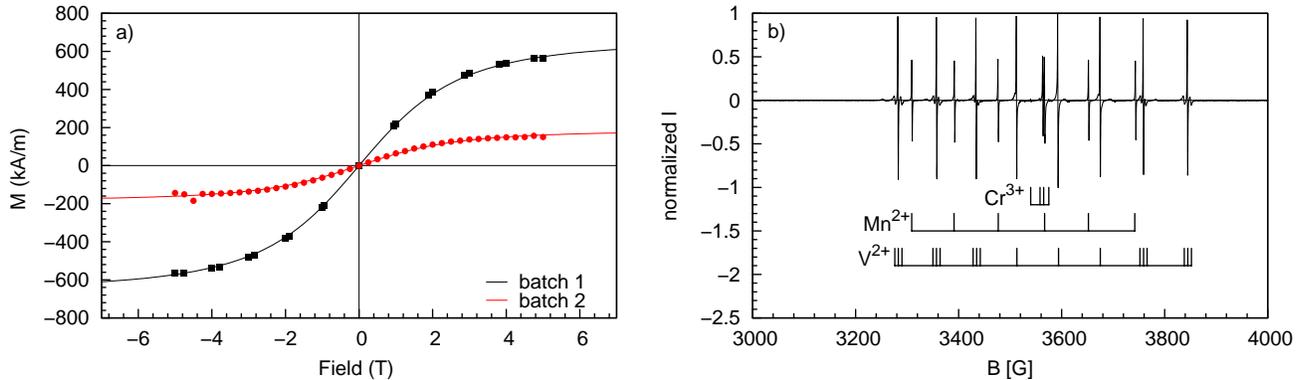}
\caption[]{a) Uncorrected magnetic hysteresis of two MgO substrates from different
 batches measured at 4~K. The spectra is fitted to a Brillouin function for the
Cr$^{3+}$, Mn$^{2+}$ and V$^{2+}$ impurities. b) A typical room temperature
epr spectrum is shown of a MgO single crystal substrate containing Cr$^{3+}$,
Mn$^{2+}$ and V$^{2+}$ impurities. } \label{fig:figure_7}
\end{figure*}

\begin{figure}
    \includegraphics[width=\columnwidth]{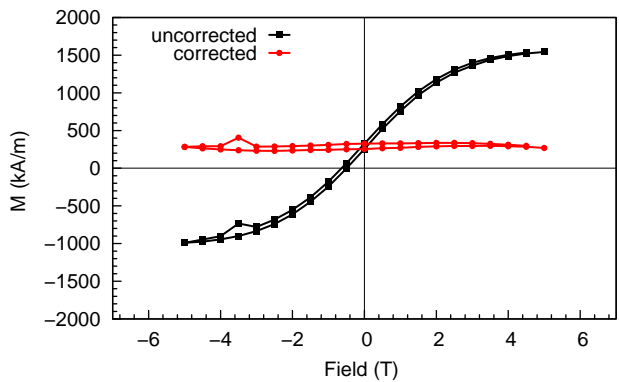}\\  
\caption{Uncorrected and corrected magnetic hysteresis of the Sm-Co film grown at 2.3~$\cdot$~10$^{-3}$~mbar at 4.2K.}
\label{fig:figure_8}%
\end{figure}

We also investigated the low temperature behavior of the Sm-Co films. In Figure~\ref{fig:figure_6}, 
the uncorrected magnetization hysteresis loops for a Sm-Co film grown at 10.5~$\cdot$~10$^{-3}$~
mbar are shown as a function of temperature. For temperatures down to 30~K, the
coercive field and the saturation magnetization increase slowly. Also, the
diamagnetic contribution of the substrate becomes smaller. When going to even
lower temperatures, the hysteresis loop shows a clear paramagnetic behavior.
Figure~\ref{fig:figure_7}(a) shows the uncorrected hysteresis loop for two MgO
substrates from different batches at 4~K. Also here the low temperature
magnetization measurements show a clear paramagnetic behavior. Electron
paramagnetic resonance (EPR) measurements were done to identify the
paramagnetic impurities. Figure~\ref{fig:figure_7}(b) shows the room temperature EPR
measurements of a MgO substrate. Clearly visible are the resonance lines of
Cr$^{3+}$, V$^{2+}$ and Mn$^{2+}$ impurities \cite{vanwieringen}.
In Figure~\ref{fig:figure_8}, the uncorrected and corrected magnetization hysteresis loops at 4.2~K for 
a Sm-Co film grown at 2.3~$\cdot$~10$^{-3}$~mbar are shown. Clearly visible is that the uncorrected 
loop is a minor loop, where the remanent magnetization for both the positive and negative field sweep has the same high and positive value. 
The magnetization hysteresis loop is corrected by subtracting the magnetization of the impurities. The corrected hysteresis 
loop is almost field independent, showing that the coercive field is larger than 5~T.

\section{Discussion}
Summarizing the experimental findings, we see that with increasing sputter
pressure, the lattice parameters and the values for $H_c$ increase until the
pressure of 6.0~$\cdot$~10$^{-3}$~mbar is reached, where the lattice parameter
corresponds to the Sm$_2$Co$_7$ alloy. At this pressure $M_s$ increases
sharply. Increasing the pressure further, $M_s$ comes down again, while $H_c$
now starts to decrease. In the whole range, we find the Sm content of the films
higher than expected from the stoichiometric ratio's of the line compounds.
Apparently, the sputter process does not result in a well-defined composition.
In particular, the films are not a mixture of the SmCo$_5$ and Sm$_2$Co$_7$
compounds, which would result in two lines with different weight in the x-ray
data. Rather, the films consist of one main composition which is
able to incorporate a varying amount of Co-atoms, probably in a complex
with the detected surplus of Sm. Cross sectional transmission electron
microscope data \cite{Tamm} have shown that when the Sm-Co film is not grown
under the optimal conditions, different epitaxial growth modes exist. The film
does not grow with a full crystalline structure and amorphous areas are formed
where the remaining material is stored.

The surface morphology in Figure \ref{fig:figure_3} shows that the grain size and
grain shape changes when the sputter pressure is changed. At high
pressures, relatively large rectangular grains are grown. When reducing the
pressure, the grains become smaller and more square-like. We surmise this
is due to the change in average energy of the atoms bombarding the substrate:
at low pressure, this energy  is higher and as a result, more defects are
created during the growth of the first Sm-Co layers. When the number of
defects becomes larger, also the number of preferred nucleation sites
increases. The increase in the number of nucleation sites results in a reduced
grain size and a more rough surface.
Magnetically, the picture is somewhat complicated by the paramagnetic
behavior of the MgO substrates, which is due to transition metal impurities, in
particular Mn, V, and Cr.  Their amount varies from batch to batch. The low
temperature magnetic hysteresis shows a clear paramagnetic behavior
superimposed on it, so we assume that the magnetization $M_{\text{imp}}$ of these
impurities can be described by the Brillouin function

\begin{equation}\label{eq:brillouin}
    \begin{split}
    M_{\text{imp}} &= NgJ \mu_B  \frac{2J+1}{2J} \coth \left( \frac{2J+1}{2J} \frac{g J \mu_B JB}{k_bT} \right)  \\
      & \quad - NgJ \mu_B \frac{1}{2J} \coth \left( \frac{1}{2J}\frac{g J \mu_B JB}{k_bT}  \right)
    \end{split}
\end{equation}
with $N$ the number of atoms, $g$ the g-factor, $\mu_B$ the Bohr magneton, $J$
the total angular momentum, $k_B$ Boltmans constant and $T$ the temperature.
The total magnetization for the three different impurities is then modelled as
the sum of the magnetization of each type of impurities.

From fitting equation~\ref{eq:brillouin} to the low temperature magnetic
hysteresis of a bare MgO substrate, we estimate that in the MgO substrates in
Figure~\ref{fig:figure_7}(a), the concentration of the impurities is in the
range of 15-60~ppm \cite{Prucnal201270}. Another factor that influences the
corrections made to the magnetization is the oxidation of the Cr capping layer,
which induces an unknown extra magnetization.

Magnetically, the properties of the films grown at different sputter background
pressures up to 6.0~$\cdot$~10$^{-3}$~mbar are very similar. At 300~K, the films
show a square hysteresis loop for the field applied in the plane of the sample,
with a coercive fields of about 3~T and a saturation magnetization of 0.4~T.
The films grown with a sputter background pressure above 6.0$\cdot 10^{-3}$
mbar show a decrease in coercive field, although the Sm concentration, lattice
constant and saturation magnetization do not change significant with respect to
the films grown at pressures below 6.0~$\cdot$~10$^{-3}$ mbar. We attribute the
change in coercive field to the larger grain size of the grown Sm-Co film. The
grain size of thin Sm-Co films is hard to control, but the grain size of
SmCo$_5$ powder can be controlled well by grinding and milling. In experiments
with nanoparticles with a different size, it was found that the size has a significant influence
on the coercive field and a optimum particle size is approximately 100-200 nm
\cite{chen:012504}.

Only the film grown at 6.0~$\cdot$~10$^{-3}$~mbar shows a large saturation
magnetization of 0.87~T. SmCo$_5$ and Sm$_2$Co$_7$ crystals have a saturation
magnetization $M_s$ of 1.05 and 1.29 T respectively. Due to the four-fold
symmetry of the MgO(100) substrate, the saturation magnetization of the films
will not have this values because part of the Sm-Co grains are aligned in-plane
and the other grains are aligned out-of-plane and do not contribute to the
saturation magnetization when measured along the (100)- or (010)-axis.

The optimal sputter pressure to grow the SmCo$_5$-like phase with a larger
saturation magnetization might be lower than 1.5$\cdot$~10$^{-3}$~mbar, the
lowest pressure at which we can grow films. This appears also from the fact
that the Sm$_2$Co$_7$-like phase can be grown over a large pressure window with
a large coercive field but only at a very small pressure window gives a high
saturation magnetization, as is shown in Figure~\ref{fig:figure_5}.

At 4.2 K, the result is a little bit more complicated. When decreasing the
temperature, the MgO substrate develops an induced moment. The Mn, Cr and V
impurities add at low temperatures a paramagnetic contribution. The variation in the magnetic moment is
quite large, as shown in Figure~\ref{fig:figure_7}(a). Furthermore, Figure \ref{fig:figure_8} shows clearly that
the measured magnetic hysteresis loop is not the hysteresis loop itself, but a minor loop, since the
coercive field is larger than the 5~T which can be reached in the magnetometer. When the contribution of the impurities in the MgO 
substrate is subtracted using equation \ref{eq:brillouin}, the Sm-Co film shows an almost constant magnetization.

A problem of growing Sm-Co films from an alloy target is, that the film
composition is very sensitive to the target composition. We used a number 
of different commercially obtained targets which, according to the vendor, 
where fabricated from the same batch of alloy material. All targets had a slightly 
different Sm concentration. From Figure \ref{fig:figure_2}, it is clear that a small change
in Sm concentration can change the magnetic and crystallographic properties a
lot. The films grown with the other targets showed coercive fields between 1.0
and 2.0 T. Further investigations with cosputtering, confirmed this assumption
that a very small change in composition has a huge influence on the grown
films. \cite{}

To conclude, by varying the sputter Ar pressure, we can change the phase from a
SmCo$_5$-like to the Sm$_2$Co$_7$-like phase. We find that the film composition is extremely 
sensitive to small variation in target composition and the Ar gas pressure 
used for sputtering. The grown films have good crystal
texture and magnetic properties.  But, the type of film is not as well defined
as might be expected.

\section{Acknowledgments}
We thank Roger W\"{o}rdenweber and Eugen Hollmann for the RBS measurements. Financial support from the European Commission 
(FP7-ICT-FET No. 225955 STELE) is gratefully acknowledged.

\end{document}